\documentclass[aps,twocolumn,showpacs,showkeys]{revtex4}
\usepackage{graphicx}
\usepackage{times}

\newcommand{\Gamnas}{Ga$_{1-x}$Mn$_x$As}

\begin{document}

\title{Memory function formalism approach to electrical conductivity and optical response of
dilute magnetic semiconductors}

\author{F. V. Kyrychenko}
\author{C. A. Ullrich}
\affiliation{Department of Physics and Astronomy, University of Missouri, Columbia, Missouri 65211}

\date{April 02, 2007}

\begin{abstract}
A combination of the memory function formalism and time-dependent density-functional theory is applied to transport in
dilute magnetic semiconductors. The approach considers spin and charge disorder and
electron-electron interaction on an equal footing. Within the weak disorder limit and using a simple parabolic approximation for
the valence band we show that Coulomb and exchange scattering contributions to the resistivity in Ga$_{1-x}$Mn$_x$As are of the
same order of magnitude. The positional correlations of defects
result in a significant increase of Coulomb scattering, while the suppression of localized
spin fluctuations in the ferromagnetic phase contributes substantially to the experimentally observed drop of resistivity below
$T_c$. A proper treatment of dynamical screening and collective excitations is
essential for an accurate description of infrared absorption.



\end{abstract}

\maketitle

%\section{Introduction}

The idea of utilizing the carrier spins in new electronic devices provides the basis for the
rapidly developing field of spintronics \cite{spintronic}.
A unique combination of magnetic and semiconducting properties makes dilute magnetic semiconductors
(DMSs) attractive for various spintronics applications \cite{ohno}. A lot of attention is
drawn to Ga$_{1-x}$Mn$_x$As since the discovery of its relatively high ferromagnetic transition
temperature \cite{ohno}, with a current record of $T_c = 159$ K \cite{record}.

The sensitivity of magnetic and transport properties of Ga$_{1-x}$Mn$_x$As to the details of growth conditions \cite{shimizu} and
post-growth annealing \cite{hayashi,potashnik,yu} points to the crucial role played by the defects and their configuration, and
has stimulated intense research on the structure of defects and their influence on the various properties of the system
\cite{timm_rev}. Most theoretical calculations of the electrical conductivity and optical response in these materials, however,
treat disorder within the relaxation time approximation, where the relaxation time is often treated as an adjustable
phenomenological parameter. It is essential, therefore, to develop a theory of electrical conductivity in DMSs with emphasis
given to disorder.

Our approach is based on a combination of the memory function
formalism \cite{Gotze81,belitz,UV} and time-dependent density-functional theory (TDDFT) \cite{TDDFT},
which not only goes beyond the simple relaxation time approximation for disorder scattering, but
allows one to consider key features of DMSs such as spin and charge disorder
and electron-electron interaction on an equal footing. The present paper is thus an extension of our previous work
\cite{Kyrychenko2007}, in which we first introduced the memory function formalism for transport in DMS,
but did not address spin disorder and electronic many-body effects beyond static screening.

%\section{General theory} \label{sec:general}
Disorder in our model consists of the Coulomb potential of the charge defects and fluctuations of localized spins (the mean-field
part of $p$-$d$ exchange interaction enters the clean system Hamiltonian). Introducing a four-component notation, one can express
the disorder Hamiltonian as $\hat{H}_I = V^2 \sum_{\bf k} \hat{\vec{\cal U}}({\bf k})\cdot \hat{\vec{\rho}}({\bf -k})$, where the
four-component disorder potential
\begin{equation}\label{pot}
  \hat{\vec{{\cal U}}}({\bf k})=\frac1{V}\sum_j \left(\begin{array}{c}
    U_j({\bf k}) \\
    \frac{J}2 \hat{S}_j^- \\
    \frac{J}2 \hat{S}_j^+ \\
    \frac{J}2 \left(\hat{S}_j^z-\langle S \rangle\right) \
  \end{array} \right) e^{i{\bf k\cdot R}_j}
\end{equation}
is coupled to the four-component charge and spin density operator $\hat{\rho}^{\mu}({\bf k})=\frac1{V}\sum_{\bf q} \sum_{\tau
\tau'} (\sigma^{\mu})_{\tau \tau'} \;
  \hat{a}^+_{{\bf q-k},\tau} \, \hat{a}_{{\bf q},\tau'} \:$.
Here, $\sigma^\mu$ ($\mu=1,+,-,z$) is defined via the Pauli matrices, where $\sigma^1$ is the $2\times 2$ unit matrix,
$\sigma^\pm=(\sigma^x \pm i \sigma^y)/2$, and the sum in (\ref{pot}) runs over all defects.

To describe transport in DMSs we employ the memory function formalism \cite{Gotze81,belitz,UV}. The central point
of this approach is the calculation of the current relaxation kernel (or memory function), whose imaginary part can be associated
with the Drude relaxation rate. Our derivation of the memory function in spin- and charge-disordered media is based on an
equation of motion approach for the current-current response function \cite{Gotze72,GV}. Technical details of the derivation will
be published elsewhere. Here we only give the final expression for the memory function in the long-wavelength limit in
paramagnetic state:
\begin{equation}\label{finalS}
M(\omega)  = \sum_{ {\bf k}\atop \mu\nu } k_{\alpha} k_{\beta}
  \left\langle\hat{{\cal U}}_{\mu}({\bf -k})\;
  \hat{{\cal U}}_{\nu}({\bf k})\right\rangle_{H_m} \widetilde{\chi}_{\rho^{\mu}\rho^{\nu}}({\bf k},\omega),
\end{equation}
where
\begin{equation}\label{chis}
  \widetilde{\chi}_{\rho^{\mu}\rho^{\nu}}({\bf k},\omega)=
  \frac{V^2}{n m \omega}\Big(\chi_{\rho^{\mu}\rho^{\nu}}({\bf k},\omega)-\chi^c_{\rho^{\mu} \rho^{\nu}}({\bf k},0) \Big),
\end{equation}
$\alpha,\beta=x,y,z$ \cite{footnote}, $n$ is the carrier concentration, averaging is performed over the magnetic subsystem
Hamiltonian $\hat{H}_m$, and $\chi_{\rho^{\mu} \rho^{\nu}}({\bf k},\omega)$ are charge- and spin-density response functions
associated with the operators $\hat{\rho}^{\mu}$. Eq.~(\ref{finalS}) contains the set of charge- and spin-density response
functions of the disordered system and, strictly speaking, should be evaluated using a self-consistent procedure. This approach
was realized in Ref. \cite{Gold} to study a spin-independent system close to the metal-insulator transition. In our case,
however, we assume that the disorder is weak enough so we can approximate $\chi_{\rho^{\mu}\rho^{\nu}}({\bf k},\omega)$ by their
clean (disorder-free) system counterparts $\chi^c_{\rho^{\mu}\rho^{\nu}}({\bf k},\omega)$.
%For simplicity we neglect the time evolution of localized spin operators, thus assuming that the carriers move through an
%ensemble of frozen spins.

The form of the disorder potential (\ref{pot}) allows us to separate in Eq.~(\ref{finalS}) contributions from charge and spin
disorder $M(\omega)=\tau_n^{-1}+\tau_s^{-1}$ with
\begin{equation}\label{taun}
\frac1{\tau_n(\omega)}  = \sum_{ \bf k} k_{\alpha} k_{\beta}
  \left|\hat{{\cal U}}_1({\bf k})\right|^2
  \widetilde{\chi}_{\rho^1\rho^1}({\bf k},\omega),
\end{equation}
and
\begin{equation}\label{taus}
 \frac1{\tau_s(\omega)} = \sum_{ \bf k \atop \mu\nu=+,-,z } k_{\alpha} k_{\beta}
  \left\langle\hat{{\cal U}}_{\mu}({\bf -k})\;
  \hat{{\cal U}}_{\nu}({\bf k})\right\rangle_{H_m}
  \widetilde{\chi}_{\rho^{\mu}\rho^{\nu}}({\bf k},\omega).
\end{equation}
The imaginary parts of Eqs.~(\ref{taun}) and (\ref{taus}) represent the charge and spin relaxation rates, which depend on
frequency and, in general, on momentum. Calculations of the charge and spin relaxation rates have been performed for the case of
$\rm Ga_{0.95}Mn_{0.05} As$, assuming for simplicity parabolic dispersion for holes. The material parameters used are: heavy hole
effective mass $m=0.5\, m_0$, dielectric constant $\varepsilon=13$, and exchange constant $VJ=55\:{\rm meV\,nm^3}$, which
corresponds to the widely used DMS p-d exchange constant $N_0 \beta=1.2\,$eV \cite{dietl}.

\begin{figure}
\centering
\includegraphics[width=1.0\linewidth]{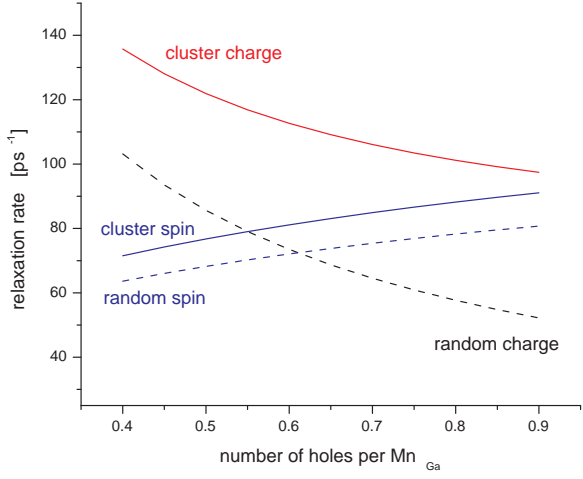}
{\caption{Charge and spin relaxation rates for random and correlated impurities in $\rm Ga_{0.95}Mn_{0.05} As$. See text for
details.} \label{fig1}}
\end{figure}

%\section{Correlated defects} \label{sec:position}
Most theoretical models for transport in DMSs assume random defect distributions. Monte-Carlo simulations of Timm {\em et al.}
\cite{timm}, however, have shown that, driven by the Coulomb attraction, donor and acceptor defects in {\Gamnas}  tend to form
clusters. The main effect of such a clustering is {\em ionic} screening of the disorder Coulomb potential, which has been shown
to be necessary to correctly reproduce the band gap, metal-insulator transition and shape of the magnetization curve.
But the correlation of defect positions also gives rise to a momentum dependent impurity structure factor. We here
account for both effects.

We consider systems with two types of defects, manganese ions in substitutional
($\rm Mn_{Ga}$) and interstitial ($\rm Mn_I$) positions. We treat the former as the acceptor centers that carry localized spins,
and the latter as spinless double donors. The donor-acceptor cross term in Eq.~(\ref{taun})
accounts for ionic screening.

Correlation in the defects positions gives rise to a set of impurity structure factors in
the product of the charge components of the disorder potential in Eq.~(\ref{taun}):
\begin{equation}
  \hat{{\cal U}}^i_1({\bf k})\; \hat{{\cal U}}^{i'}_1({\bf -k})=
  U^i_1({\bf k}) U^{i'}_1({\bf -k})\frac{n_i}V \:{\cal S}_{ii'}({\bf k}).
\end{equation}
Here, the structure factors
can be expressed through the corresponding pair distribution functions $P_{ii'}$,
\begin{equation}
 {\cal S}_{ii'}({\bf k})=\delta_{ii'}+\frac{n_{i'} V}{\Omega_0}\int_V P_{ii'}(R) \cos({\bf k\cdot R}) d{\bf R},
\end{equation}
where $i,i'$ label acceptors and donors and $\Omega_0$ is the elementary cell volume. We use a simple model expression for
$P_{ii'}$ (for details, see \cite{Kyrychenko2007}), with parameters consistent with the Monte Carlo simulations of
Ref.~\cite{timm}.

Positional correlations alone, however, are not sufficient to affect the spin relaxation rate (\ref{taus}), orientational
correlations of the spin fluctuations are also necessary. To account for these, we considered interacting spins described by the
Heisenberg Hamiltonian $\hat{H}_m=-\frac12 \sum_{j \neq j'} J_{jj'}\; \hat{\bf S}_j \cdot \hat{\bf S}_{j'}$ and use a high
temperature expansion to obtain the following expression for the product of the spin components of the disorder potential in
Eq.~(\ref{taus}):
\begin{equation}
\left\langle\hat{{\cal U}}_{\mu}({\bf k})\; \hat{{\cal U}}_{\nu}({\bf -k})\right\rangle_{H_m}=
  \delta_{\mu \nu} \frac{J^2}4 \frac{S_{\rm Mn}(S_{\rm Mn}+1)}3
\frac{n_i}V \:{\cal S}_{s}({\bf k}),
\end{equation}
where $\mu,\nu=x,y,z$ and the spin structure factor ${\cal S}_s({\bf k})$ (adjusted for orientational correlations) is
\begin{equation}
 {\cal S}_s({\bf k})=1+\frac{2S_{\rm Mn}(S_{\rm Mn}+1)}{3N_i}
  \sum_{j>j'} \frac{J_{jj'}({\bf R}_{jj'})}{k_B T}\cos\left({\bf k\cdot R}_{jj'}\right).
\end{equation}
Here, $S_{\rm Mn}=5/2$ and $N_i$ are the spin and number of localized moments, and the $d-d$ exchange constants
$J_{jj'}$ are chosen to reproduce the typical experimental value of $T_c=150$ K within the standard mean field approach.

In Fig.~\ref{fig1} we plot the static ($\omega=0$) relaxation rates (\ref{taun})-(\ref{taus}) calculated for random (dashed
lines) and correlated (solid lines) impurities in $\rm Ga_{0.95}Mn_{0.05} As$ as a function of
the level of compensation. Correlated
impurities were modelled by clusters containing 10 $\rm Mn_{\rm Ga}$ with an average concentration
$x_c=0.1$ of substitutional
Mn ions within the cluster. It is seen that the charge and spin contributions to the relaxation rate are of the similar
magnitude. The combined effect of ionic screening and impurity structure factor
results in a net increase of the charge relaxation rate for correlated impurities for the whole range of compensations. The increase is
significant (up to 100\%) and is sensitive to the cluster configurations. The latter might be controlled by the post growth
annealing. The positional correlation of the scattering centers also leads to an increase of the spin relaxation
rate for interacting spins at room temperature.
This effect, however, is smaller than that for charge relaxation.

\begin{figure}
\centering
\includegraphics[width=1.0\linewidth]{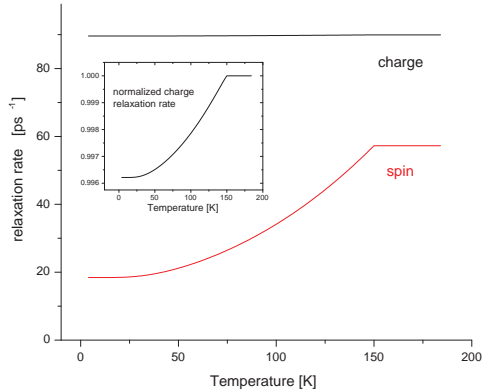}
\caption{Temperature dependence of the charge and spin relaxation rates for $\rm Ga_{0.95}Mn_{0.05} As$ with $T_c=150$ K. Inset:
charge relaxation rate, normalized to its value in the paramagnetic state.} \label{fig2}
\end{figure}

%\section{Magnetic ordering}\label{sec:order}
Magnetic ordering is known to have a strong effect on the transport properties of DMSs \cite{omiya}. The resistivity of optimally
annealed samples reveals a pronounced maximum at critical temperature and decreases significantly for temperatures below $T_c$
\cite{spintronic1}. Lopez-Sancho and Brey \cite{lopez} proposed to explain the resistivity change in terms of the variation of
the Fermi surface and the transport scattering time when going from the paramagnetic to the ferromagnetic phase. Their model,
however, completely neglects the scattering off the fluctuations of localized spins. On the other hand, spin fluctuations are
effectively suppressed in the ferromagnetic state. In the fully spin-polarized state scattering takes place only due to the
quantum fluctuations of localized spins.

In Fig. \ref{fig2} we present the temperature dependence of spin and charge relaxation rates calculated according to
Eqs.~(\ref{taun}) and (\ref{taus}) for $\rm Ga_{0.95}Mn_{0.05} As$ with $T_c=150$ K. Suppression of localized spins fluctuations
below $T_c$ leads to a significant reduction of the spin relaxation rate. Given the comparable magnitudes of charge and spin
relaxations, the 70\% drop in the latter translates into about 20\% reduction in total resistivity,
which is consistent with
both experimental observations and the calculations of Ref.~\cite{lopez}. This shows that the spin scattering is clearly not
negligible in \Gamnas, especially if one considers effects associated with magnetic ordering.

We also determined the variation of the charge scattering rate with temperature,
which was discussed in Ref.~\cite{lopez}. However, we find its
magnitude to be much smaller, see the inset in Fig.~(\ref{fig2}). This is most likely due to the fact that in our
calculations we used a simple parabolic band and isotropic spin splitting. Nevertheless, it is clear that
further work is required to adequately describe
both scattering mechanisms and their contributions to the experimentally observed drop in resistivity.

%\section{Electron-electron interaction} \label{sec:interaction}
Most previous studies of (magneto)transport in DMS included electronic many-body effects only in the form of static Coulomb
screening \cite{sinova,jungwirth,rates}. However, this simplification ignores the role of dynamical many-body effects such as the
coupling to plasmon modes. A major advantage of the memory function formalism is that it allows to consider both disorder and
electron-electron interaction on equal footing. All carrier many-body effects in Eq.~(\ref{finalS}) including screening,
correlations and collective excitations are absorbed in the set of density and spin-density response functions and, in principle,
can be accounted for exactly by means of TDDFT \cite{TDDFT}. In the original work by
Gross and Kohn \cite{grosskohn}, the interacting density-density response function of a homogeneous system is shown to be
representable as
\begin{equation} \label{eq.5}
\chi^{-1}({\bf q},\omega) = \chi_0^{-1}({\bf q},\omega) - \frac{4\pi}{q^2} - f_{\rm xc}({\bf q},\omega) \;,
\end{equation}
with the exchange-correlation kernels $f_{\rm xc}$.

\begin{figure}
\centering
\includegraphics[width=0.95\linewidth]{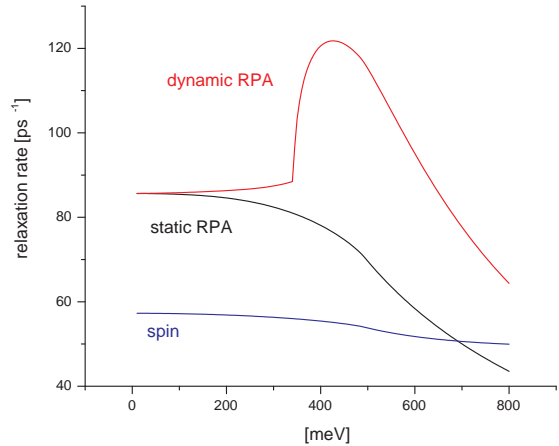}
\caption{Frequency dependence of the charge and spin relaxation rates for $\rm Ga_{0.95}Mn_{0.05} As$ with electron-electron
interaction taken into account within static and dynamic RPA.} \label{fig3}
\end{figure}

Coupling to the charge plasmon mode already occurs on the level of dynamic RPA, corresponding to the first two terms on right
hand side of Eq.~(\ref{eq.5}) generalized for the multicomponent spin-charge density response functions. Fig.~(\ref{fig3}) shows
the frequency dependence of charge and spin relaxation rates calculated for $\rm Ga_{0.95}Mn_{0.05} As$ within static and dynamic
RPA. Coupling to plasmon modes results in a strong enhancement of the charge relaxation rate since it provides an efficient
channel to absorb the momentum from impurity scattering.

This approximation, however, does not affect the spin relaxation. To capture collective spin modes one has to go beyond RPA
and include exchange and correlation contributions in Eq.~(\ref{eq.5}), which can be done in
the adiabatic local-density approximation \cite{ullrichflatte}.  However, our single band model
does not produce any spin collective mode. In a
a more realistic model with multiple valence bands, inter-valence band spin
collective modes may play a role. This is currently work in progress.

%\section{Conclusions} \label{sec:conclusion}
To summarize, we have derived a general framework for combining the memory function formalism with TDDFT in spin and charge
disordered media, to study transport properties of DMSs. Within the weak disorder limit and using a simple parabolic model for
the valence band we have shown that Coulomb and exchange scattering contributions to the  resistivity in Ga$_{1-x}$Mn$_x$As are
of the same order of magnitude and should be taken into account simultaneously. The combined effect of ionic screening and
impurity structure factor results in a net increase of the relaxation rate in systems with positional correlation of the defects.
The suppression of localized spin fluctuations in ferromagnetic phase contributes substantially to the drop of resistivity
experimentally observed below $T_c$. Our calculations further suggest that the effects of collective electron dynamics on
transport and optical conductivity in DMS is significant. One can expect that there will be a distinct collective signature in
the mid-infrared free-carrier absorption, which has not been included in most previous theoretical studies. These results, in
particular when validated in currently ongoing work which includes band structure details, should give valuable insight into the
transport properties of DMSs.

\acknowledgments This work was supported by DOE Grant No. DE-FG02-05ER46213.


\begin{thebibliography}{99}
\bibitem{footnote} In general, the memory function is a tensor
$M_{\alpha\beta}$. For macroscopically isotropic systems in the long-wavelength limit, however, its three nonvanishing components
are identical: $M_{xx}=M_{yy}=M_{zz}=M(\omega)$.
\end{thebibliography}
\end{document}